\documentclass[% aip, jmp, bmf, sd, rsi, amsmath,amssymb, %preprint,%
 reprint,author-year, %author-numerical, Conference Proceedings
]{revtex4-1}
\usepackage{graphicx}% Include figure files
\usepackage{dcolumn}% Align table columns on decimal point
\usepackage{bm}% bold math
\usepackage{float}
\usepackage[utf8]{inputenc}
\usepackage[T1]{fontenc}
\usepackage{mathptmx}
\usepackage{etoolbox}
\makeatletter
\def\@email#1#2{%
 \endgroup
 \patchcmd{\titleblock@produce}
  {\frontmatter@RRAPformat}
  {\frontmatter@RRAPformat{\produce@RRAP{*#1\href{mailto:#2}{#2}}}\frontmatter@RRAPformat}
  {}{}
}%
\makeatother
\begin{document}

\preprint{AIP/123-QED}

\title[Sample title]{Molecular Mechanisms Underlying the Effects of Urea and the Structural Dynamics of Bovine Serum Albumin}

% Force line breaks with \\
\author{Yanis R. Espinosa}
%\altaffiliation[Also at ]{Universidad de Pamplona, Pamplona, Colombia, Facultad de Ciencias Básicas, Grupo de Investigación en Biología Molecular y Genética.}
\affiliation{University of Pamplona, Faculty of Basic Sciences, Research Group in Molecular Biology and Genetics (BIOMOGEN), Pamplona, Colombia.}
\affiliation{Institute of Physics of Liquids and Biological Systems (IFLYSIB), CONICET, La Plata, Argentina.}

\author{C. Manuel Carlevaro}%
%\email{Second.Author@institution.edu.}
\affiliation{Institute of Physics of Liquids and Biological Systems (IFLYSIB), CONICET, Argentina.}
\affiliation{Universidad Tecnológica Nacional, Facultad Regional La Plata, Grupo de Materiales Granulares, La Plata, Buenos Aires, Argentina.}

\author{C. Gastón Ferrara}
%\homepage{https://mel.iflysib.unlp.edu.ar/}
\email{gastonf@iflysib.unlp.edu.ar}
\affiliation{Institute of Engineering and Agronomy, National University Arturo Jauretche, Av Calchaqui no.~6200, Florencio Varela, Argentina.}
\affiliation{Institute of Physics of Liquids and Biological Systems (IFLYSIB), CONICET, Argentina.}

%Second institution and/or address%\\This line break forced% with \\}%

\date{\today}% It is always \today, today, %  but any date may be explicitly specified

\begin{abstract}
The disruption of protein structures by denaturants such as urea is well-documented, although the underlying molecular mechanisms are not yet fully understood. In this study, we employed Molecular Dynamics (MD) simulations to examine the effects of urea on the structural stability of Bovine Serum Albumin (BSA) at concentrations ranging from 0 M to 5 M. Our results reveal that urea induces a dehydration–rehydration cycle by displacing and partially substituting water molecules in BSA’s hydration shell. At lower concentrations, urea decreases protein–water hydrogen bonding while enhancing protein–urea interactions. At higher concentrations, urea tends to aggregate, which limits direct interactions with the protein, promotes rehydration, and leads to alterations in the tertiary structure, although the secondary structure remains largely preserved. These findings offer mechanistic insights into urea-induced protein denaturation and stability.
\end{abstract}

\maketitle
%%%%%%%%%%%%%%%%%%%%%%%%%%%%%%%%%%%%%%%%%%%%%%%%%%%%%%%%%%%%%%%%%%%%%%%%%%%%%%%%%%
%%%%%%%%%%%%%%%%%%%%%%%%%%%%%%%%%%%%%%%%%%%%%%%%%%%%%%%%%%%%%%%%%%%%%%%%%%%%%%%%%%%%%%%%%%%%%%%%%
\section{\label{sec:level1}Introduction} 

The study of urea and other denaturing agents in aqueous protein solutions has been a central topic in biochemistry for decades. Although the ability of these compounds to disrupt protein tertiary structure is well documented, the precise molecular mechanisms underlying their action remain a subject of ongoing debate \cite{wallqvist1998hydrophobic, frank1968structural, aastrand1991properties, rezus2006effect, soper2003impact, gahtori2023role, stumpe2007interaction, lindgren2010stability, moeser2014unified, matubayasi2018all, nnyigide2018exploring, kumaran2011denaturation, ma2014microscopic, sinibaldi2008sans, monhemi2014protein}.

Multiple mechanisms have been proposed to explain urea-induced protein denaturation. Two predominant models are traditionally recognized: the \textbf{direct mechanism}, in which urea disrupts intramolecular stabilizing forces—such as hydrogen bonds and van der Waals interactions—by directly interacting with the protein; and the \textbf{indirect mechanism}, in which urea alters the structure and dynamics of the solvent, thereby weakening hydrophobic interactions that stabilize the protein's native conformation \cite{rossky2008protein, das2009urea, niether2018unravelling, hua2008urea, jha2014kinetic, guckeisen2021effect}. Recent studies, however, suggest that these mechanisms are not mutually exclusive but may act synergistically, with their relative contributions depending on specific experimental conditions  \cite{khan2019protein, paladino2022structure, lim2009urea}.

The impact of urea on protein stability depends on multiple factors, including denaturant concentration, temperature, pH, and thermodynamic parameters \cite{miller2022urea}. Comparative studies with guanidinium chloride have revealed distinct conformational effects on proteins \cite{baptista2023, candotti2013, maity2020}. Computational evidence indicates that urea exhibits preferential affinity for hydrophobic residues, promoting solvation and disrupting hydrophobic collapse \cite{lapidus2007protein}, ultimately compromising native structural integrity \cite{candotti2013}.

Recent molecular dynamics simulations have shown that under acidic conditions, urea tends to reorient locally near protonated acidic residues (Asp/Glu), primarily acting as a hydrogen bond acceptor \cite{de2020shift}. Notably, these local rearrangements have minimal impact on urea’s interactions with the protein backbone or nonpolar residues, supporting the hypothesis of a largely pH-independent denaturation mechanism.Complementing these computational findings, surface-sensitive vibrational spectroscopy has experimentally confirmed urea’s direct interactions with hydrophobic side chains at protein interfaces \cite{gahtori2023role}. These interactions lower the hydrophobic barrier, enhance interfacial water structuring, and disrupt side chain organization, thereby destabilizing the protein’s hydrophobic core.
Additionally, molecular dynamics (MD) simulations have shown that the dipolar orientation of urea in aqueous solution can temporarily stabilize certain secondary structural elements before their complete disruption \cite{baptista2023}. This finding suggests that denaturation is not an instantaneous event but a dynamic process governed by the competition between protein-protein and protein-solvent interactions \cite{paladino2023, gooran2024}.

In this study, we employ MD simulations to investigate the effects of varying urea concentrations on bovine serum albumin (BSA) structure. This protein was selected due to its ability to undergo reversible conformational changes in response to pH variations \cite{scanavachi2020aggregation}. Specifically, we focus on the N-isoform of BSA at pH 6.5, a physiologically stable and functional conformation. Our results exhibit strong agreement with both experimental and previous computational studies, reinforcing the validity of our approach and contributing to a deeper understanding of urea-induced protein denaturation.

%%%%%%%%%%%%%%%%%%%%%%%%%%%%%%%%%%%%%%%%%%%%%%%%%%%%%%%%%%%%%%
%%%%%%%%%%%%%%%%%%%%%%%%%%%%%%%%%%%%%%%%%%%%%%%%%%%%%%%%%%%%%%
\section{Materials and methods: Systems and simulations}

We investigated the impact of urea on the structural integrity of bovine serum albumin (BSA) in aqueous solutions across varying urea concentrations. The computational model utilized for urea was the Boek model \cite{boek1994interfaces} with nonbonded interactions and covalent parameters sourced from the GROMOS 54A7 force field \cite{schmid2011definition}. For water, we employed the Simple Point Charge Extended (SPC/E) model \cite{berendsen1987missing}, as developed by Berendsen and colleagues. The initial atomic coordinates for the BSA model were obtained from the crystal structure of bovine serum albumin deposited in the Protein Data Bank (PDB code 4F5S) \cite{bujacz2012structures}. The topology of BSA at a standard pH of 6.5 was obtained from previous work and validated through Small-Angle X-ray Scattering (SAXS) and MD simulations \cite{scanavachi2020aggregation}. The net charge of the protein under these conditions was +11, and sodium (Na$^{+}$) counterions were added to neutralize the system. 

Intermolecular interactions were explicitly accounted for using the GROMACS 2020.6 molecular dynamics simulation package \cite{spoelgromacs}. As a reference, a control system devoid of urea was first simulated, with water molecules randomly distributed around the protein (see Control System in Table 1).

%%%%%%%%%%%%%%%%%%%%%%%%%%%%%%%%%%%%%%%%%%%%%%%%%%

\begin{table}[H]
\centering

\begin{tabular}{ccccc}
\hline
\textbf{Systems} & \textbf{\begin{tabular}[c]{@{}c@{}}Water \\ molecules\end{tabular}} & \textbf{\begin{tabular}[c]{@{}c@{}}Contraions \\ Na$^{+}$\end{tabular}} & \textbf{\begin{tabular}[c]{@{}c@{}}Urea \\ molecules\end{tabular}} & \textbf{\begin{tabular}[c]{@{}c@{}}Urea \\ concentrations\end{tabular}} \\ \hline
Control               & 61,212                                                              & 11                                                                      & 0                                                                  & 0 M                                                                     \\
1M               & 57,312                                                              & 11                                                                      & 1,099                                                              & 1 M                                                                     \\
2M               & 53,356                                                              & 11                                                                      & 2,198                                                              & 2 M                                                                     \\
3M               & 49,235                                                              & 11                                                                      & 3,278                                                              & 3 M                                                                     \\
4M               & 44,840                                                              & 11                                                                      & 4,396                                                              & 4 M                                                                     \\
5M               & 40,963                                                              & 11                                                                      & 5,468                                                              & 5 M                                                                     \\ \hline
\end{tabular}
\label{table1}
\caption{The simulated systems. Bovine Serum Albumin (BSA) in water (control), and BSA in urea-water solutions at varying urea concentrations ranging from 1 to 5 M.}
\end{table}

%%%%%%%%%%%%%%%%%%%%%%%%%%%%%%%%%%%%%%%%%%%%%%%%%%%
The optimization of the control system was carried out in two distinct stages. Initially, the system was equilibrated under the NVT ensemble for 500 ps, during which the C$\alpha$ atoms of the protein were restrained with a harmonic potential using a force constant of 1000 kJ mol$^{-1}$ nm$^{-2}$ applied in each Cartesian direction. This restraint allowed for the equilibration of the protein-water system, effectively minimizing any potential artifacts that could arise from the subsequent introduction of urea. In the second stage, the position restraints were removed, and the system was equilibrated under the NpT ensemble for an additional 500 ps to allow for proper density adjustment and pressure equilibration.

\begin{figure*}[ht!]
\centering\includegraphics[width=0.6\linewidth]{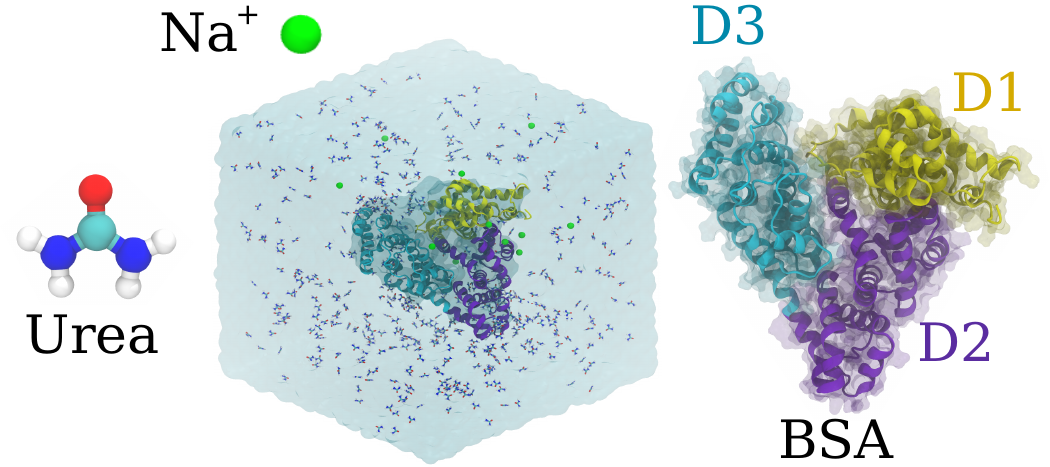}
\caption{Water box simulation featuring urea, sodium, and Bovine Serum Albumin (BSA). BSA divided into three domains: D1 (yellow, residues 1-185), D2 (violet, residues 186-378), and D3 (light blue, residues 379–576). Sodium ions (Na$^{+}$) depicted in green. Urea molecules are represented by cyan carbon, red oxygen, blue nitrogen, and white hydrogen atoms.}
\label{Fig:1}
\end{figure*}

After optimization of the control system, the final equilibrated conformation was used as the reference structure for subsequent simulations involving urea. Urea molecules were randomly inserted into this initial configuration (Figure \ref{Fig:1}), yielding five simulation boxes containing BSA solvated in water–urea mixtures at increasing concentrations ranging from 1 M to 5 M (see Table 1). Each system was subjected to the same two-step optimization protocol previously applied to the control system, ensuring methodological consistency across all conditions. Production simulations were then carried out for 500 ns. From the final 8 ns of each trajectory, eight distinct configurations were randomly selected to serve as starting points for additional 10 ns simulations. Each replicate was initiated with a different set of initial velocities to enhance statistical sampling. This protocol enabled a detailed investigation of the long-timescale dynamics and intermolecular interactions within the protein–urea–water systems.

%%%%%%%%%%%%%%%%%%%%%%%%%%%%%%%%%%%%%%%%%%%%%%%%%%%%%
%%%%%%%%%%%%%%%%%%%%%%%%%%%%%%%%%%%%%%%%%%%%%%%%%%%%%
%%%%%%%%%%%%%%%%%%%%%%%%%%%%%%%%%%%%%%%%%%%
%%%%%%%%%%%%%%%%%%%%%%%%%%%%%%%%%%%%%%%%%%%%%%%%

The optimization and production processes of both the control system and BSA in water-urea solutions were simulated using a cubic box with periodic boundary conditions, at $300.15$ K and $1$ bar of pressure, employing a velocity rescale thermostat \cite{bussi2007canonical} and Berendsen barostat \cite{berendsen1984molecular}, respectively. The coupling time constants utilized in the thermostat and barostat were $0.1$ ps and $1.0$ ps, respectively. Electrostatic interactions were computed using the PME method \cite{abraham2011optimization} with a cutoff for Van der Waals and Coulomb interactions set at $1.0$ nm. The LINCS algorithm \cite{hess1997lincs} was employed to constrain solute bonds. A time step of $2$ fs was maintained throughout the simulations.

\emph{Hydrogen Bonds}. The number of hydrogen bonds per molecule was calculated using the standard GROMACS utility \emph{gmx hbond}. A hydrogen bond was considered to exist when the donor–acceptor distance was $\leq$0.35 nm and the acceptor–donor–hydrogen angle (A–D–H) was $\leq$30$^{\circ}$, by commonly accepted geometric criteria in GROMACS.

%%%%%%%%%%%%%%%%%%%%%%%%%%%%%%%%%%%%%%%%%%%%%%%%%%%%%%%%%%%%%%%%%%%%%
%%%%%%%%%%%%%%%%%%%%%%%%%%%%%%%%%%%%%%%%%%%%%%%%%%%%%%%%%%%%%%%%%%%%%
\section{Results and Discussion}
\label{S:3}

All analyses are based on the average of eight independent simulations, each 10 ns in duration, conducted for each system studied (control without urea and urea:water mixtures from 1 to 5 M). Before data collection, all systems underwent 500 ns of hydration/solvation to ensure stabilization.

\subsection{Hydrogen Bond Analysis}

Initially, we analyzed the formation of protein-water hydrogen bonds (HBs) as a function of increasing urea concentration. For each of the eight systems corresponding to each concentration, the calculations were averaged over the final 8 ns of the simulation. Figure 1 in the supplementary material displays an example of the time evolution of the number of hydrogen bonds between the protein and water, averaged across the eight systems for the control condition.

\begin{figure}[htbp]
\centering\includegraphics[width=1.1\linewidth]{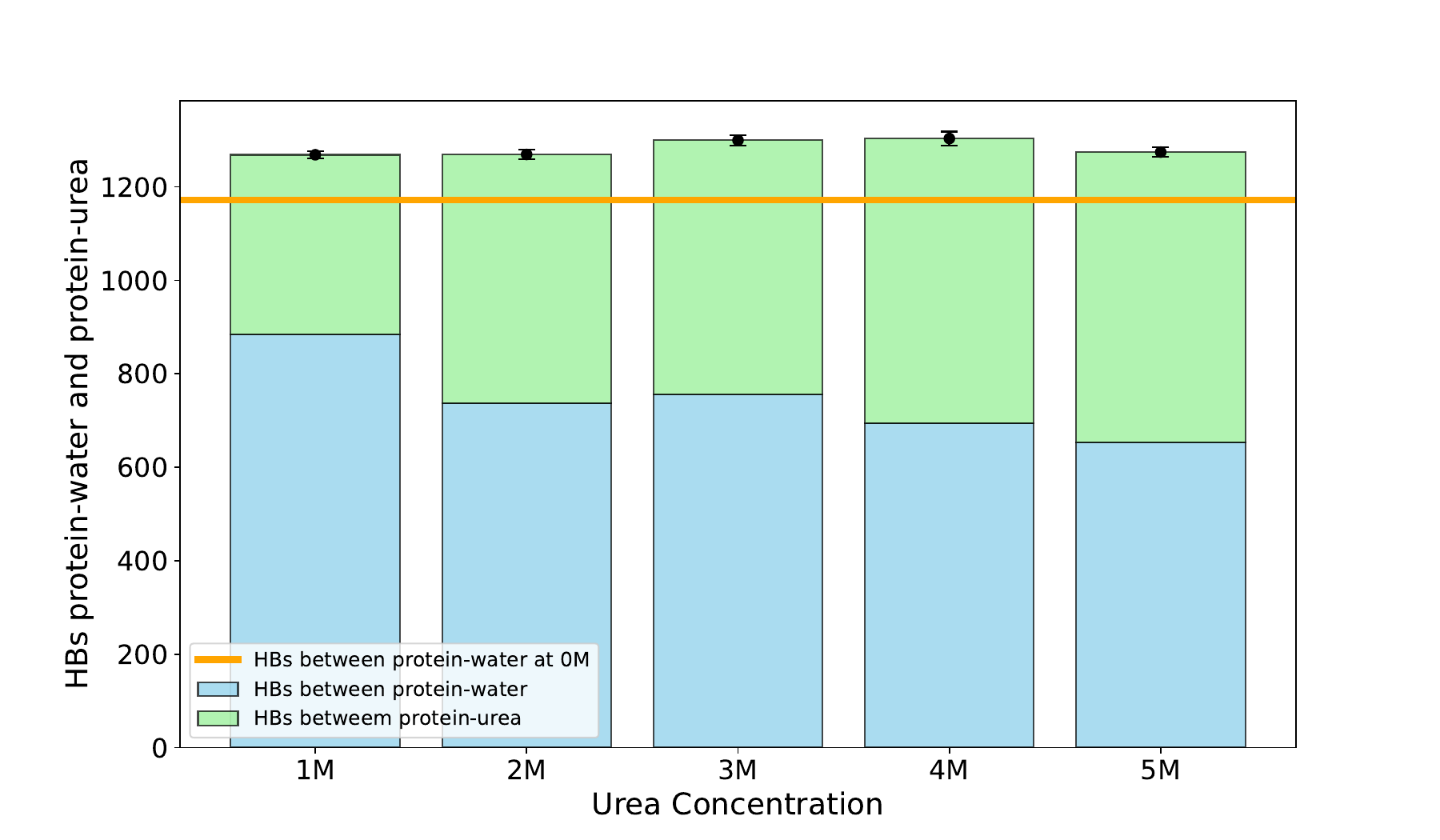}
\caption{Total number of hydrogen bonds (HBs) formed between BSA and the components of urea:water mixtures as a function of urea concentration. Light blue bars represent the average number of HBs between the protein and water molecules, while light green bars correspond to HBs formed between the protein and urea molecules. The horizontal orange line indicates the average number of protein–water HBs in the absence of urea (0 M), used as a reference. Error bars represent the standard deviation across eight replicates. The data illustrate the progressive substitution of protein–water HBs by protein–urea HBs with increasing urea concentration, while the total number of protein–solvent HBs remains relatively constant. Figures 2, 3, and 4 in supplementary material show the HBs distributions for protein–water, protein–urea, and urea–urea interactions, normalized by the number of molecules of each species.}
\label{Fig2}
\end{figure}

As shown in Figure \ref{Fig2}, the presence of urea in the BSA–water system leads to a marked reduction in protein–water HBs. At 1 M urea, a decrease of about 25\% is observed, rising to 45\% as the concentration increases from 2 M to 5 M. Interestingly, beyond this point, the number of protein–water HBs plateaus at around 700, indicating that urea-induced dehydration reaches a threshold beyond which further displacement of water molecules from the protein surface is limited.

Similarly, Figure \ref{Fig2} displays the average number of hydrogen bonds between the protein and urea molecules, highlighted in light green. Notably, the decrease in protein-water HBs with increasing urea concentration is proportional to the increase in protein-urea HBs. This observation suggests that as urea concentration rises, the protein undergoes progressive dehydration, accompanied by a corresponding increase in urea solvation.
In this case, we observe that as the concentration increases, the average number of HBs between urea molecules and the protein initially rises but eventually stabilizes around 600. This plateau suggests that both urea-induced dehydration and urea solvation may reach a threshold, beyond which additional urea molecules do not significantly alter the hydration state of the protein.

Taking the average number of protein–water HBs at 0 M urea as a reference, Figure \ref{Fig2} shows that BSA consistently maintains over 1,200 hydrogen bonds across all urea concentrations. Although increasing urea concentration induces protein dehydration, resulting in the loss of protein–water HBs, a compensatory gain in protein–urea HBs is observed. However, the overall increase in the total number of HBs formed between BSA and the urea:water binary mixtures is approximately 2\%.
This suggests that urea molecules increasingly interact with the BSA, potentially displacing water molecules and altering the hydrogen bonding network around the protein. 

In addition, as the urea concentration increases, the total number of HBs between urea molecules grows monotonically (see Figure 4a in the supplementary material). However, when normalized by the number of urea molecules in the system (see Figure 4b in the supplementary material), the number of HBs per urea molecule exhibits an asymptotic increase with concentration. Notably, beyond 3 M, this value begins to stabilize, whereas at 1 M it remains relatively low. This behavior suggests that as the concentration increases, urea molecules form a more interconnected hydrogen-bond network that gradually approaches saturation.

%%%%%%%%%%%%%%%%%%%%%%%%%%%%%%%%%%%%%%%%%%%%%%%%%%%%%%%%
%%%%%%%%%%%%%%%%%%%%%%%%%%%%%%%%%%%%%%%%%%%%%%%%%%%%%%%%%%
%%%%%%%%%%%%%%%%%%%%%%%%%%%%%%%%%%%% cambiar a partir de aca

\subsection{Radial Distribution Function (RDF)}

To support our initial findings, we analyzed the radial distribution functions (RDFs) calculated between the center of mass (COM) of the protein and that of water molecules, and separately, between the protein COM and the COM of urea molecules.

\begin{figure}[htbp]
\centering\includegraphics[width=0.9\linewidth]{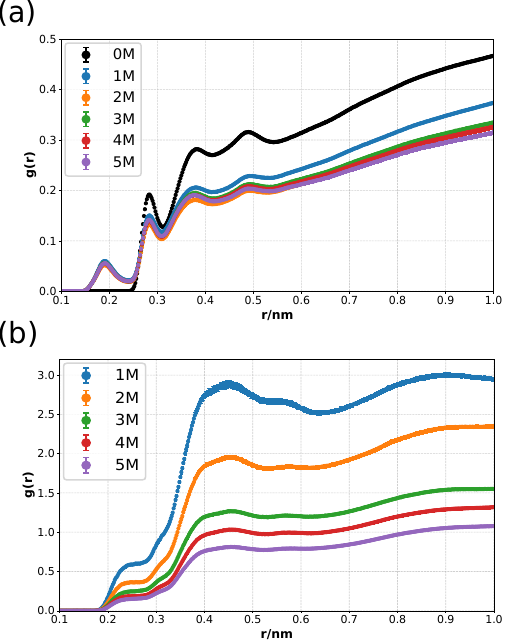}
\caption{Radial distribution functions (RDFs) calculated between the center of mass (COM) of the protein and the COM of solvent components at different urea concentrations. (a) RDF between the protein COM and the COM of water molecules for the range 0.1{-}1.0 nm. (b) RDF between the protein COM and the COM of urea molecules for the same distance range. A redistribution of solvent density near the protein is observed as urea concentration increases.} 
\label{Fig3}
\end{figure}

In Figure \ref{Fig3}(a), the RDF between the COM of the protein and that of water molecules shows that, even in the presence of urea, water molecules can approach the protein COM as closely as $\sim$0.20 nm. This suggests that local hydration near solvent-accessible regions of the protein core is maintained across all urea concentrations. Additionally, a secondary peak appears around 0.28–0.30 nm, which corresponds to the outer hydration shell. This feature is most prominent in the absence of urea (0 M) and progressively decreases in intensity with increasing urea concentration, indicating a disruption of long-range water structuring around the protein. However, a deviation from this trend is observed at 2M urea, suggesting a shift in molecular behavior. This deviation is difficult to correlate directly with the total average number of HBs between the protein and water molecules (Figure \ref{Fig2}), as the values are not normalized with respect to the number of solvent molecules. 

By analyzing the normalized average number of hydrogen bonds between protein and water (Figure 2 in the Supplementary Material) and its variation in the presence of urea, we show that the result is consistent with the observations in \ref{Fig3}(a). From this, it can be observed that from 0 M to 2 M, the normalized average number of HBs between the protein and water decreases, reaching a minimum at 2 M. Beyond this minimum, the normalized average number of HBs between the protein and water increases, accompanied by the corresponding rehydration of the protein. 
Although the number of water molecules decreases by approximately 30\% when comparing the systems at 1 M and 5 M urea concentrations (Figure \ref{Fig2}), the normalized average number of hydrogen bonds (HBs) remains nearly constant at both concentrations.

When analyzing the RDF between urea molecules and the protein, shown in Figure \ref{Fig3}(b), two distinct peaks are evident at 1 M urea. The first peak ($\sim$0.20{–}0.30 nm) corresponds to urea molecules close to the protein surface, indicating short-range interactions. The second peak ($\sim$0.40{–}0.50 nm) is associated with urea molecules located in the outer solvation shell. As the urea concentration increases, the intensity of both peaks progressively decreases, suggesting a reduced presence of urea near the protein and a possible redistribution into the bulk solvent.

The analysis based on the number of HBs between urea molecules and the protein, normalized by the number of urea molecules (Fig. 3 in the supplementary material), reinforces and complements the observations in Fig. \ref{Fig3}(b). As the urea concentration increases, the number of HBs formed with the protein remains relatively constant at high concentrations, leading to a monotonic decrease in the average normalized HB count.

We interpret the sustained HBs count at high urea concentrations as evidence that protein solvation by water molecules persists, while direct interactions with urea molecules diminish. Complementarily, we analyzed the formation of hydrogen bonds between the protein main chain, protein side chain, and water and urea molecules (see Figures 5 and 6 in the supplementary material).These results suggest that increasing urea concentration does not induce significant changes in the interactions between urea or water molecules and the protein main chain, but does markedly affect their interactions with the protein side chains. This supports the notion that urea–protein interactions are more likely to occur with the side chains rather than the backbone.

\begin{figure*}[htbp]
\centering\includegraphics[width=0.8\linewidth]{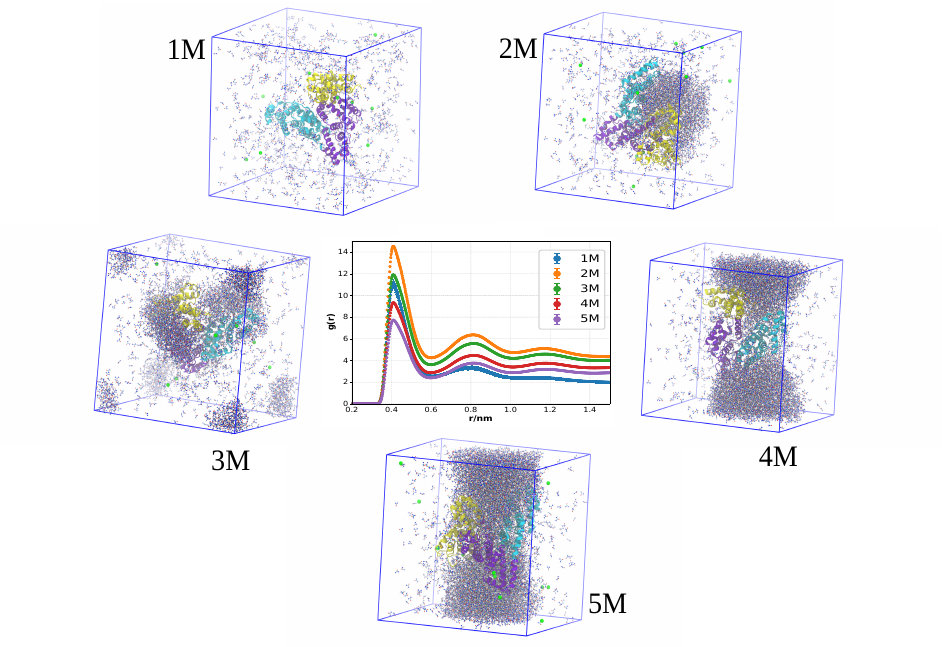}
\caption{RDF between the carbon atoms of urea molecules and different snapshots of the final configurations at each urea concentration.} 
\label{Fig4}
\end{figure*}

When examining the radial distribution function (RDF) between urea carbons to better understand urea's role in its interactions with the protein (Figure \ref{Fig4}). At lower urea concentrations (1M to 2M), there is an increase in both the central and second peaks of the RDF. At higher concentrations ($>$2M), however, a decrease in the central peak is observed, indicating a shift in behavior compared to the trends at 1M and 2 M. This decrease at concentrations above 2 M can be attributed to specific geometric arrangements that urea molecules adopt during aggregation, as shown in Figure \ref{Fig4}.

The inset images in Figure \ref{Fig4} reveal these changes in urea aggregation geometry, which are consistent with the model proposed by Susan and collaborators \cite{atahar2019aggregation}. 
The complementary analysis made with the normalized hydrogen bond distribution (see Fig. 4b in the supplementary material) among urea molecules aligns with the aggregation changes depicted in Fig. \ref{Fig4}. This suggests that as urea concentration increases, the preference for urea-urea interactions also increases, and this tendency toward self-interaction appears to become dominant at higher concentrations.

%%%%%%%%%%%%%%%%%%%%%%%%%%%%%%%%%%%%
%%%%%%%%%%%%%%%%%%%%%%%%%%%%%%%%%%%%%%%%%%%%%%%%%%%%%%%%%%%%%%%%%%%%%%%

\subsection{Radius of Gyration (RG), Root Mean Square Deviation (RMSD) and Solvent Accessible Surface Area (SASA)}

All the parameters presented in this section were analyzed over the last eight nanoseconds of the eight replica simulations performed at each urea concentration. Reported values correspond to averages computed across these datasets.

\begin{figure*}[htbp]
\centering\includegraphics[width=0.8\linewidth]{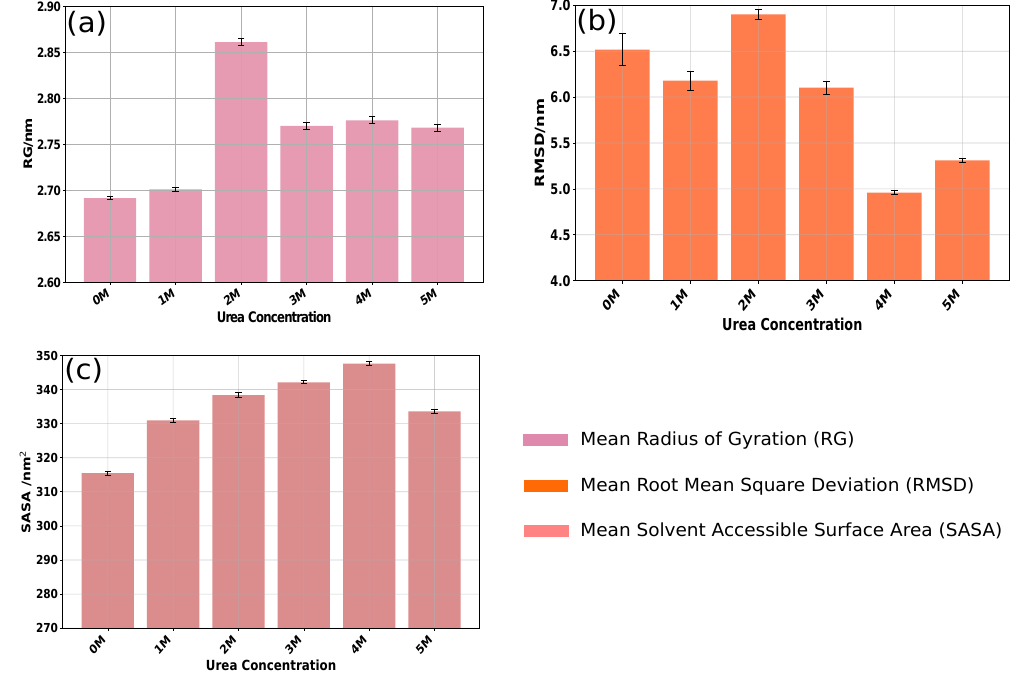}
\caption{(a) Mean Radius of gyration vs urea concentration. (b) Mean of Root Mean Square Deviation vs urea concentration. (c) Mean solvent accessible surface area of the protein vs urea concentration.}
\label{Fig5}
\end{figure*}

We analyzed the protein's radius of gyration as a function of urea concentration and observed that the average radius increases with increasing urea concentration. At 2 M, the radius of gyration increases more pronouncedly, as shown in Figure \ref{Fig5}(a). At this concentration, the interaction between the protein and water is minimal, corresponding to the lowest number of hydrogen bonds per water molecule formed between the protein and water molecules (Figure 2 in the supplementary material).
Overall, the presence of urea increases the average radius of gyration compared to the system without urea. However, these changes remain modest, not exceeding 6\%, suggesting that significant conformational alterations are unlikely. This finding is consistent with the observations of Yasuda et al. \cite{hayashi2007protein}, who reported that within this range of urea concentrations and temperatures, the protein structure remains relatively stable, with only minor perturbations driven by solvent dynamics.

The analysis of the average RMSD as a function of urea concentration, presented in Figure \ref{Fig5}(b), was performed by comparing the structures concerning the one obtained from the Protein Data Bank (PDB code: 4F5S).

Most urea concentrations reduce the average RMSD compared to the protein's native state in the absence of urea, suggesting structural stabilization. Notably, RMSD values at 1M and 3M urea are comparable, whereas significantly lower values at 4M and 5M indicate enhanced stabilization. In contrast, at 2M urea, the average RMSD increases relative to 0M, suggesting structural deviation. The destabilization observed at 2M urea correlates with changes in the protein's radius of gyration under the same conditions. These findings suggest a relationship between structural destabilization and protein expansion at this concentration. Moreover, the data presented in this figure support the RG analysis, reinforcing the conclusions regarding the role of water and its interaction with the protein.

When analyzing the mean value of the total solvent-accessible surface area ( Figure \ref{Fig5}(c)), it is evident that the introduction of urea leads to an increase in solvent exposure. In the absence of urea, the protein exhibits the lowest SASA, which increases monotonically up to a concentration of 4 M. The percentage increases in the average SASA values slightly exceed 9\% only at a urea concentration of 4 M; in all other cases, they remain below this level compared to the protein in the absence of urea.
Interestingly, at higher urea concentrations, deviations from this monotonic behavior emerge. While these deviations do not contradict the overall trend, they underscore the need for a more detailed analysis, particularly one that focuses on how the SASA changes across individual protein domains. Given the structural complexity of proteins such as BSA, a domain-specific SASA analysis is essential for understanding how different regions of the protein respond to varying urea concentrations, especially regarding solvent exposure and potential local conformational changes.

\begin{figure*}[htbp]
\centering\includegraphics[width=1.0\linewidth]{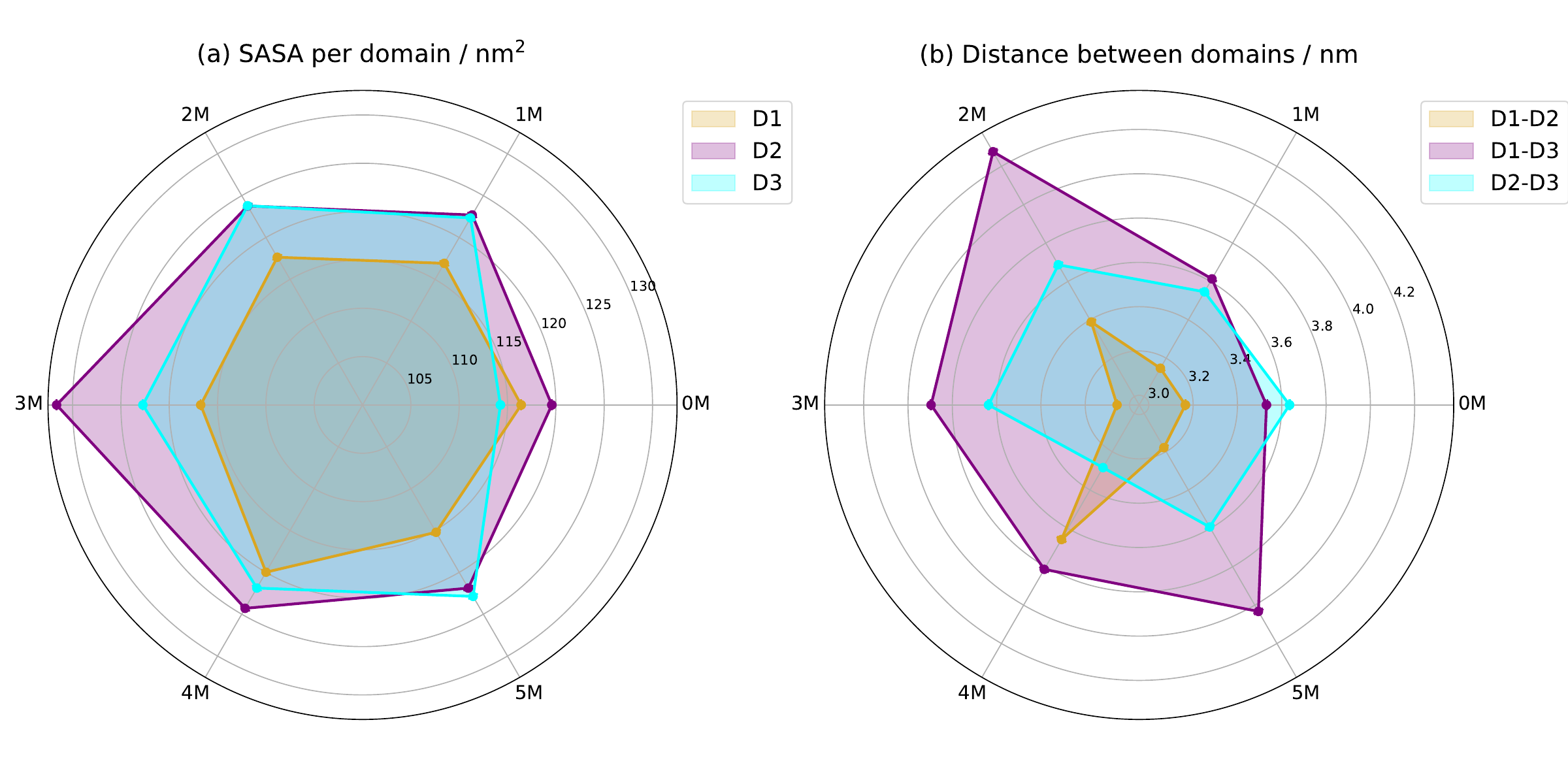}
\caption{(a) Mean solvent accessible surface area per domain vs urea concentration. (b) Mean distance between the center of mass of the domains as a function of concentration. The errors in the mean values shown in the figures are small, making them barely visible without enlarging the figure.}
\label{Fig6}
\end{figure*}

To refine the analysis, we investigated the SASA by domain and the inter-domain distances to identify the protein regions most affected by hydration changes or variations in urea concentration (Figure \ref{Fig6}).

Across all urea concentrations, Domain 1 consistently exhibits the lowest solvent-accessible surface area. In contrast, Domain 2, consistent with previous studies \cite{scanavachi2020aggregation}, demonstrates the greatest propensity for surface exposure, displaying the highest SASA values at nearly all urea concentrations. Notably, for all concentrations, the solvent-exposed surface area of all three domains increases compared to the system without urea.

The presence of urea increases the SASA of each domain, observing that the contribution at total SASA is different, and dependent on urea concentration. To better understand the role of each domain, we analyzed the interdomain distance in Figure \ref{Fig6}(b).

As an initial observation, Figure \ref{Fig6}(b) shows that at 0M urea, the distances between D1-D3 and D2-D3 are approximately equal, both near 3.6 nm. In contrast, the D1-D2 distance is around 3.2 nm, suggesting that the centers of mass of the domains form an isosceles triangle.  As the urea concentration increases, the distance between the D1 and D3 domains increases relative to the control system without urea. However, the distances between the D1-D2 and D2-D3 domains fluctuate around the reference value with the addition of urea. Interestingly, at 2 M urea, the inter-domain distances D1-D3 tend to increase, followed by a decrease at 3 M. Additionally, at 4 M and 5 M urea, the D1-D2 distance exhibits an opposite behavior compared to the D1-D3 and D2-D3 distances. This suggests that Domain 3 is the most affected by urea concentration, as it moves away from Domain 1 and 2, leading to a greater solvent-exposed surface area for this region (Figure \ref{Fig6}(a).

%%%%%%%%%%%%%%%%%%%%%%%%%%%%%%%%%%%%%%%%%%%%%%%%%%%%%%%%%%% ACA DEBO ACOMODAR
%%%%%%%%%%%%%%%%%%%%%%%%%%%%%%%%%%%%%%%%%%%%%%%%%%%%%%%%%%%%%%%%%%%%%%
\subsection{Secondary Structure}

To gain deeper insights into how urea impacts the protein structurally, we analyzed its secondary structure as a function of urea concentration.

\begin{figure*}[htbp]
\centering\includegraphics[width=0.9\linewidth]{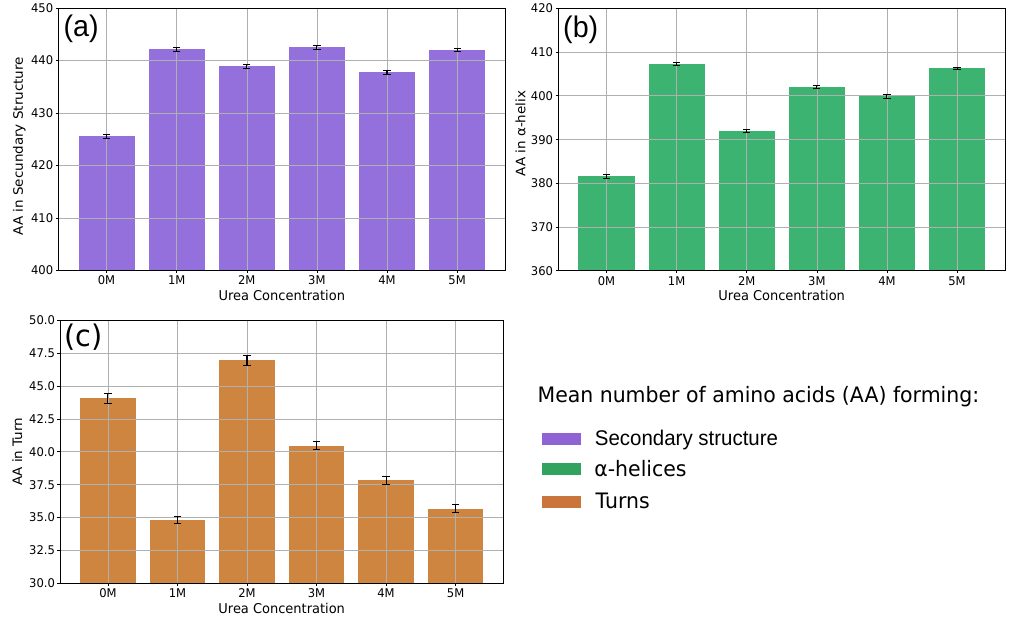}
\caption{Mean number of amino acids (AA) involved in different types of secondary structure as a function of urea concentration. (a) Total number of amino acids forming any type of secondary structure. (b) Number of amino acids forming $\alpha$-helices. (c) Number of amino acids forming turns.}
\label{Fig7}
\end{figure*}

As shown in Figure \ref{Fig7}(a), the presence of urea induces a slight increase in the number of amino acids participating in the secondary structure. This increase, however, is modest, with the largest observed difference being just over 4\% when comparing the absence of urea to 5M. Overall, urea concentration does not appear to play a significant role in driving changes to the secondary structure. The variations in the number of amino acids involved in the secondary structure between 1M and 5M are minimal, amounting to slightly more than 1\%.

A more detailed secondary structure analysis reveals a clear trend of increasing $\alpha$-helix formation with urea concentration (Figure \ref{Fig7}(b)). At all concentrations, the number of $\alpha$-helices is consistently higher than at 0M, with the largest increase observed at 1M. Beyond 2M, the number of amino acids forming $\alpha$-helices increases almost monotonically with concentration.

In contrast, Figure \ref{Fig7}(c) shows that the average number of amino acids forming turns is generally reduced compared to 0 M, except at 2 M, where it peaks above the initial value. Beyond 2 M, the number of turns decreases monotonically with increasing urea concentration.

These findings suggest urea reinforces the secondary structure by promoting $\alpha$-helix formation. However, this is accompanied by a general, albeit smaller, reduction in the number of amino acids forming turns. As a preliminary hypothesis, it can be proposed that the structural changes induced by urea primarily affect the protein's tertiary structure. The observed difference of approximately 3-4\% in the average number of amino acids in the secondary structure corresponds to around fifteen residues. This numerical difference implies that the loss of hydrogen bonds observed in the protein is not directly associated with reductions in its secondary structure.

\section{Conclusion}

This study employed molecular dynamics (MD) simulations to investigate the effects of urea on the structural behavior of bovine serum albumin (BSA) at physiological pH. Across the tested urea concentrations (0-5 M), our results demonstrate that urea primarily disrupts the tertiary structure, while the secondary structure-particularly the $\alpha$-helical content-remains largely preserved. This structural resilience is consistent with previous experimental findings and highlights the stability of BSA’s N-isoform under mildly denaturing conditions \cite{murayama2004heat, abrosimova2016ftir}.

A key observation is the progressive dehydration of the protein surface with increasing urea concentration, driven by the replacement of protein–water hydrogen bonds with protein–urea interactions. Despite this shift, the total number of hydrogen bonds involving the protein remains nearly constant, suggesting a dynamic solvation equilibrium.

Analysis of radial distribution functions (RDF) and solvation metrics-including solvent-accessible surface area (SASA) and radius of gyration (Rg)-reveals modest protein expansion, particularly at intermediate urea concentrations. Domain-specific differences in solvent accessibility were observed, with the most pronounced changes occurring in Domain 2. This expansion correlates with subtle structural rearrangements, which may represent early stages of tertiary structure unfolding.

At higher concentrations, urea-urea interactions become dominant, saturating protein-urea contacts and promoting partial rehydration of the protein. This saturation occurs when all accessible surface sites on the protein are occupied, preventing further urea binding. As a result, excess urea molecules self-associate to form structured aggregates rather than interacting with the protein \cite{atahar2019aggregation}. The system thus reaches a dynamic equilibrium, wherein protein-urea interactions plateau despite further increases in urea concentration. These findings highlight a concentration-dependent balance between dehydration, solvation, and domain-specific flexibility, supporting the notion that urea-induced denaturation progresses gradually and selectively affects distinct aspects of protein structure.

Although the limited number of Na$^{+}$ ions in our system constrains the evaluation of their role in protein destabilization, future studies should investigate how urea modulates electrostatic interactions under more physiologically relevant ionic conditions. Previous studies on SDS micellar systems have shown that urea can disrupt structural organization by solvating counterions and weakening electrostatic stabilization \cite{espinosa2018mechanisms}. A similar mechanism may be operative in protein systems, particularly in cases where ion–protein interactions play a significant role in maintaining structural integrity. Exploring these effects could yield novel insights into the interplay between ion binding, solvation dynamics, and urea-induced protein denaturation.

%\begin{acknowledgments}
\section*{ACKNOWLEDGMENTS}
We wish to thank and fondly remember Dr. Raúl Grigera, who taught and shaped us in this beautiful profession. We also thank the anonymous reviewers for their thoughtful feedback on earlier drafts of this article. This work was supported by grants from the Consejo Nacional de Investigaciones Científicas y Técnicas (KE3-11220210100918CO) and the National University Arturo Jauretche (80020230100029UJ).

The Supplementary Material accompanying this study includes detailed methodological descriptions, auxiliary computations, and extended analyses that support the findings and conclusions presented in the main manuscript. Specifically, it contains an analysis of the equilibrium conditions of the studied systems and an evaluation of the number of hydrogen bonds formed between the protein and water or urea molecules, along with their respective normalizations. In addition, it reports the number of hydrogen bonds between urea molecules and their normalization. Finally, it presents an analysis of the average formation of hydrogen bonds involving the protein main chain, protein side chain, and water molecules. The material is provided in a structured and accessible format to facilitate reference.

\section*{AUTHOR DECLARATIONS}

\subsection*{Conflict of Interest}
The authors have no conflicts to disclose.

\subsection*{Ethics Approval}

Ethics approval is not required.

\section*{DATA AVAILBILITY}
The data that support the findings of this study are available from the corresponding author upon reasonable request.

%\section*{REFERENCES}\sout{lower than at 0M}
% If you have acknowledgments,
%\end{DATA AVAILBILITY}
%%%%%%%%%%%%%%%%%%%%%%%%%%%%%%%%%%%%%%%%%%%%%%%%%%%%%%%%%
%%%%%%%%%%%%%%%%%%%%%%%%%%%%%%%%%%%%%%%%%%%%%%%%%%%%%%%%%%%%%%%%
%\nocite{*}
\bibliography{Molecular-mechanism-underlying}%
%\bibliography{aipsamp}% Produces the bibliography via BibTeX.
\end{document}